# Single Top Quark Production as a Probe for Anomalous Moments at Hadron Colliders[*]

THOMAS G. RIZZO

*Stanford Linear Accelerator Center*

*Stanford University, Stanford, CA 94309*

## Abstract

Single production of top quarks at hadron colliders via $gW$ fusion is examined as a probe of possible anomalous chromomagnetic and/or chromoelectric moment type couplings between the top and gluons. We find that this channel is far less sensitive to the existence of anomalous couplings of this kind than is the usual production of top pairs by $gg$ or $q\bar{q}$ fusion. This result is found to hold at both the Tevatron as well as the LHC although somewhat greater sensitivity for anomalous couplings in this channel is found at the higher energy machine.

Submitted to Physical Review **D**.

[*]Work supported by the Department of Energy, contract DE-AC03-76SF00515.

The discovery of the top quark at the Tevatron by both the CDF and D0 Collaborations [1, 2] has renewed interest in what may be learned from a detailed study of top properties. One point of view is that this clear discovery of the top represents a great triumph and confirmation of the predictions of the Standard Model(SM), in that the top lies in the mass range anticipated by precision electroweak data[3]. Another viewpoint is that the subtleties of top quark physics itself may shed some light on new physics beyond the SM. Indeed, due to its large mass, it is widely believed that top quark physics will be the first place where non-standard effects will appear.

If the top does have non-SM interactions associated with a new mass scale it may be possible to express them in the form of higher dimensional non-renormalizable operators. These are naturally divided into those associated with the strong interactions, *i.e.*, QCD, and those associated with the electroweak sector. New interactions for the top quark in both sectors have been discussed in the literature[4, 5] and may arise as a result of, *e.g.*, compositeness or new dynamics associated with fermion mass generation[6]. In the case of QCD, the lowest dimensional operator representing new physics and conserving $CP$ that we can introduce is the anomalous chromomagnetic moment $\kappa$. On the otherhand, the corresponding chromoelectric moment, $\tilde{\kappa}$, violates $CP$. In this modified version of QCD for the top the $t\bar{t}g$ interaction Lagrangian takes the form

$$\mathcal{L} = g_s \bar{t} T_a \left( \gamma_\mu + \frac{i}{2m_t} \sigma_{\mu\nu}(\kappa - i\tilde{\kappa}\gamma_5) q^\nu \right) t G_a^\mu, \qquad (1)$$

where $g_s$ is the strong coupling constant, $m_t$ is the top quark mass, $T_a$ are the color generators, $G_a^\mu$ is the gluon field and $q$ is the outgoing gluon momentum. (Due to the non-Abelian nature of QCD, a four-point $t\bar{t}gg$ interaction is also generated, but this will not concern us in the present work.)

The study of the tree-level effect of a non-zero value of $\kappa$ at high energy $e^+e^-$ colliders,

such as the NLC, requires a high precision examination of the tail of the gluon jet energy spectrum in the process $e^+e^- \to t\bar{t}g$. Although the sensitivity to non-zero values of $\kappa$ and/or $\tilde{\kappa}$ is quite high[4] in this process, such an analysis is unfortunately many years away and so we must turn our attention to what can be learned at hadron colliders. The pair production of top via $q\bar{q}, gg \to t\bar{t}$ at both the Tevatron and LHC in the case of non-vanishing anomalous couplings has already been considered[4]. It was found that both the LHC and, eventually, the Tevatron are sensitive to values of $\kappa$ of order 0.1. This was demonstrated in detail in our earlier work for the Tevatron and will be summarized below for the LHC for purposes of comparison. Present cross section measurements at the Tevatron being performed by CDF and D0 are probing values of $\kappa$ and $\tilde{\kappa}$ which are somewhat larger, of order 0.2-0.3. It thus seems natural to ask if this potential new physics is accessible through any other top quark production channels at hadron colliders.

In the present analysis, we turn our attention to what may be learned about $\kappa$ and $\tilde{\kappa}$ through an examination of single top production via $gW \to t\bar{b}$[7]. We anticipate that this production mechanism is far less sensitive to these anomalous couplings than is the usual pair production process. The reason for this is abundantly clear: the cross section receives its dominant contribution from the $u-$channel $b$ quark exchange diagram which has no anomalous $t\bar{t}g$ vertex associated with it. To see if our expectations are indeed realized and to complete the analysis of the influence of anomalous couplings on top production we proceed with the calculation. In the SM, assuming $m_t = 175$ GeV, single top production at both the Tevatron and LHC occurs with a cross section only a factor of $\simeq 5$ or so less than that for top pairs, thus implying that adequate statistics should eventually be available at either machine to probe for anomalous effects in this channel. To show the rather weak dependence of this process on the values of $\kappa$ and $\tilde{\kappa}$, we will make use of the Effective Gauge Boson Approximation(EGBA) [8] to greatly simplify our calculations. We find that the cross section estimates obtained in this manner are sufficient for our purposes since the contributions due to anomalous couplings are so weak.

The relevant subprocess to examine for sensitivity to $\kappa$ and $\tilde{\kappa}$ in single top production is $g(q) + W(k)^+ \to t(p_t) + \bar{b}(p_b)$ (+ h.c.), which includes the $gt\bar{t}$ vertex in the diagram with t-channel top exchange. Denoting the $W$ polarization vector by $\epsilon$ and for the moment neglecting the mass of the $b$-quark, i.e., $m_b = 0$, we obtain the following parton level differential cross section

$$\frac{d\sigma}{dz} = \frac{G_F M_W^2 \alpha_s(m_t)}{24\sqrt{2}s} \frac{2p_t}{\sqrt{s}} [T_1 + \kappa T_2 + (\kappa^2 + \tilde{\kappa}^2) T_3] \tag{2}$$

where $z \equiv \cos\theta^*$, with $\theta^*$ being the top quark production angle in the center of mass frame, and $p_t$ is the magnitude of the top quark three-momentum. The $T_i$ are given by

$$\begin{aligned}
T_1 &= \frac{2}{ut'^2} \left[ t'(u^2 + t'^2) + 4(t'^2 + 2um_t^2)\epsilon \cdot q \epsilon \cdot p_b + 4ut'\epsilon \cdot q\epsilon \cdot p_t - 4t'^2(\epsilon \cdot p_b)^2 \right. \\
&\quad \left. -4ut'(\epsilon \cdot p_t)^2 + 4[t'(s - m_t^2) - 2um_t^2]\epsilon \cdot p_b \epsilon \cdot p_t \right] , \\
T_2 &= \frac{2}{ut'} \left[ u^2 - 2(s - m_t^2)(\epsilon \cdot q)^2 - 2\epsilon \cdot q[(2u + t')\epsilon \cdot p_b + u\epsilon \cdot p_t] \right] , \\
T_3 &= \frac{1}{2m_t^2} \left[ s - m_t^2 + 4\epsilon \cdot p_b \epsilon \cdot p_t \right] ,
\end{aligned} \tag{3}$$

where $s$, $u$, and $t' = t - m_W^2$ are the usual sub-process Mandelstam variables. In our numerical analysis, we will keep $m_b$ finite and evaluate $\alpha_s$ at the scale $\mu = m_t$, which we perform via 3-loop renormalization group equation evolution from $\alpha_s(M_Z) = 0.125$[3]. Since $m_t$ is not far from $M_Z$, this procedure is not greatly influenced by the use of 3-loop evolution. However, since the higher order QCD corrections to this process have not yet been calculated there is still reasonable sensitivity to the choice of $\mu$.

This sub-process cross section is apparently sensitive to the nature of the polariza-tion of the incoming $W$. To obtain the full cross section, we first sum over the weighted contributions of the longitudinal and transverse $W$'s for a given incoming quark flavor and then sum over the weighted quark densities. To be specific, we use the Martin, Roberts and Stirling MRSA and MRSA' parton densities[9], since they are in very good agreement with the latest data from the Tevatron and HERA. This particular choice of parton densities does not affect our results in any substantive manner. We assume that the scattering takes place in the $x - z$ plane with the incoming $W$ and $g$ three-momenta along the $z-$axis. In this case, we can choose the three $W$ polarization states, $\epsilon_T^i (i = x, y)$ and $\epsilon_L$ so that $\epsilon_T^i \cdot q = 0$ and $\epsilon_L \cdot q = (s - M_W^2)/(2s)$. We also obtain the following explicit expressions for the other dot products in Eq. (3):

$$\begin{aligned}
\epsilon_T^1 \cdot p_b &= -\epsilon_T^1 \cdot p_t = -p_b(1-z^2)^{\frac{1}{2}}, \\
\epsilon_T^2 \cdot p_b &= \epsilon_T^2 \cdot p_t = 0, \\
\epsilon_L \cdot p_b &= (p_W E_b - E_W p_b z)/M_W, \\
\epsilon_L \cdot p_t &= (p_W E_t + E_W p_t z)/M_W,
\end{aligned} \qquad (4)$$

where $p_i$ and $E_i$ are the magnitude of the momenta and energies of the various particles in the parton frame. Similarly,

$$\begin{aligned}
t &= -2E_g(E_t - p_t z) + m_t^2, \\
u &= -2E_g(E_b + p_b z) + m_b^2.
\end{aligned} \qquad (5)$$

From the kinematics it is easily seen that any terms in the cross section which are linear in $\tilde{\kappa}$ must vanish, as they should, since the total cross section is not a $CP$-violating observable.

In order to compare the sensitivity of the top pair and single production modes to non-zero anomalous couplings via cross section measurements at the Tevatron and LHC, we must determine how well these cross sections can be determined from future data. This issue has been a subject of extensive study by a large number of groups, the most complete and extensive on the experimental issues being that performed by the Top Quark Working Group at the TeV2000 Workshop[10, 11] and we will use their preliminary results in our analysis below. This working group considered how well the pair and single top cross section can be measured as a function of the Tevatron integrated luminosity, accounting for uncertainties due to statistics, machine luminosity, tagging efficiencies, lepton and jet acceptances, and backgrounds from other processes. For the pair production process, the estimated cross section uncertainty was found to be 13(9, 5, 4, 3.5)% for $\mathcal{L}$ =1(2, 10, 25, 100)$fb^{-1}$. At a luminosity of $\mathcal{L}$ =1 $fb^{-1}$, the error sources were 8.4% from acceptance, 10% from backgrounds, 3.5% from the machine luminosity uncertainty and the remaining due to statistics. For single tops, the corresponding uncertainties were found to be 14(10, 5, 4, 3.5)%, respectively. Essentially, the growing statistics associated with the ever-increasing integrated luminosity allows for dramatic reductions in both the systematic as well as statistical errors. At very high luminosities, the largest remaining uncertainty is the machine luminosity itself, a situation that will also be dramatically realised at the LHC with $\mathcal{L}$ =100 $fb^{-1}$.

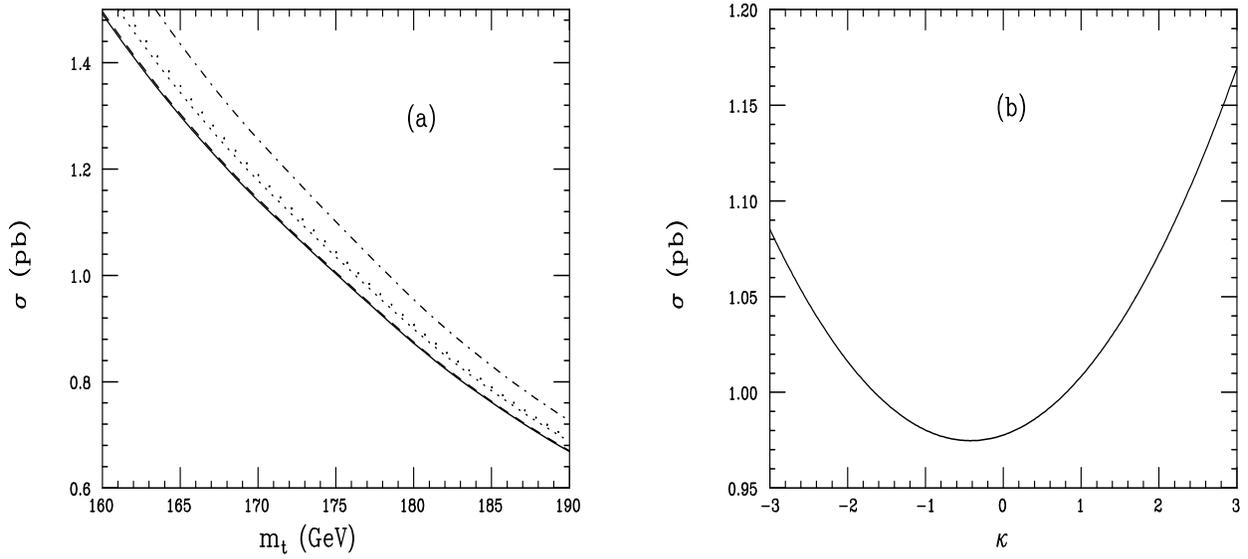

Figure 1: (a) Cross section for the process $gW^+ \to t\bar{b}(+ h.c.)$ as a function of $m_t$ at the Tevatron. The solid curve is the SM prediction whereas the dashdot(solid dot,dot, dash) curve corresponds to $\kappa = 2(-2, 1, -1)$. MRSA parton densities are assumed. (b) $\kappa$ dependencies of the cross section shown in (a) for $m_t = 175$ GeV. In both plots, $\tilde{\kappa} = 0$ is assumed.

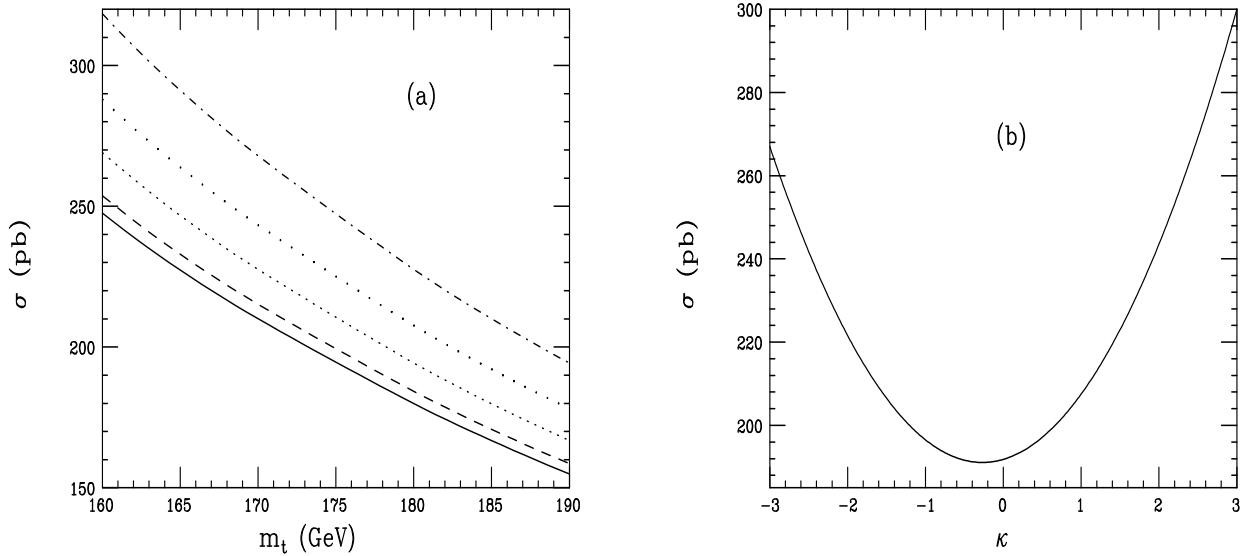

Figure 2: Same as Fig.1 but for the LHC.

On the theoretical side, top pair production at the Tevatron is now a very well studied process with full NLO calculations, including gluon resummation, now available[12]. The present uncertainties, from the Berger and Contopanagos analysis, are dominated by the choice of scale($\simeq$ 10%), parton densities($\simeq$ 5%), and the as yet uncalculated full NNLO contributions, which are expected to be small. Given the rapid evolution in this area, we can expect the total theoretical error to be at or below the 10% level by the end of the decade. In the case of single tops, the theoretical error remains rather large at present, $\simeq$ 30%. In particular, only the tree level result is presently available for the $gW \rightarrow t\bar{b}$ subprocess of interest to us[13]. It seems likely, however, that this situation will substantially improve over the next few years, particularly after single top production is directly observed at the Tevatron and the $t\bar{t}$ channel is well understood. We thus might expect that the theoretical uncertainty in the cross section for single top production may eventually drop to a level comparable to that obtainable for top pairs.

Let us first consider the case where $\tilde{\kappa} = 0$. Fig.1 shows both the dependence of the total cross section on $m_t$ for several values of $\kappa$, as well as the $\kappa$ dependence of the cross section for $m_t$ fixed to 175 GeV at the Tevatron. We note two features immediately: (i) a non-zero value for $\kappa$ almost always leads to a cross section increase except for the case of very small negative values and (ii) the difference between the SM result and that with $\kappa$ of order 2 is only of order 10%! Thus a determination of the cross section with a combined theoretical and experimental error of about 10% centered on or near the SM prediction would tell us only that $-2.9 \leq \kappa \leq 2.1$. (This 10% value is probably the best that can be done based on the discussion above and we will use it as a suggestive figure for purposes of comparison.) A similar study of the $\kappa$ dependence of the $gg$, $q\bar{q} \to t\bar{t}$ would yield sensitivities about a factor of 20 or so better as we showed in our previous work[4]. This difference is due to the lack of sensitivity in the parton-level cross section itself and cannot be overcome by better statistics, of which there is always more in the pair production channel. Of course, as the average parton center of mass energy increases and the top becomes relatively light, i.e., $m_t^2/\hat{s} << 1$, the sensitivity to $\kappa$ increases both due to the growing importance of the t-channel exchange as well as the different momentum dependence in the anomalous coupling term in the interaction Lagrangian. Thus in Fig.2, which shows the corresponding cross section results for the LHC, we see that there is an enhanced dependency on $\kappa$. A 10% determination (i.e., combined theoretical and experimental errors) centered on the SM value would restrict the range of $\kappa$ to $-1.6 \leq \kappa \leq 1.1$. Although this is an improvement it cannot match the pair production mode at either the Tevatron or the LHC for sensitivity. To verify this claim, we show the $\kappa$ dependence of the top pair production cross section at the LHC's $\kappa$ dependence in Fig.3. Here we directly see that an overall uncertainty of 10% in the cross section allows us to probe $\kappa$ values of order 0.1 or less.

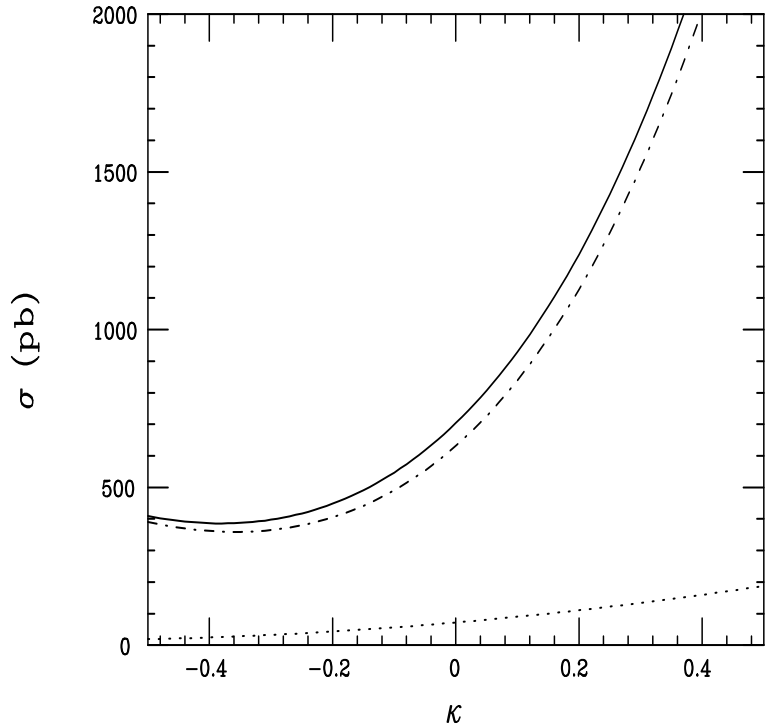

Figure 3: Cross section for $t\bar{t}$ production as a function of $\kappa$ at the LHC for $m_t = 180$ GeV. The dotted(dash-dotted) curve is the $q\bar{q}(gg)$ contribution and the solid line is their sum. MRSA' parton densities were assumed.

What happens in the reverse case, i.e., when $\tilde{\kappa}$ only is non-zero? Since $\tilde{\kappa}$ appears only quadratically in the cross section, we can restrict ourselves to semi-positive definite values of this parameter. The resulting cross sections for the Tevatron and LHC are shown in Figs. 4 and 5, respectively. The general features are quite similar to the $\kappa$ case in that non-zero values of $\tilde{\kappa}$ increase the cross section and the magnitude of the effect is comparable to that with non-vanishing $\kappa$. Here, a 10% determination centered on the SM value would yield $\tilde{\kappa} \leq 2.5$ at the Tevatron and $\leq 1.4$ at the LHC, respectively. We thus conclude that to probe for either anomalous chromomagnetic or chromoelectric moment couplings of top to gluons, the cross section in the single production channel can in no way compete with that for pair production due to greatly reduced sensitivity even when large statistics is available.

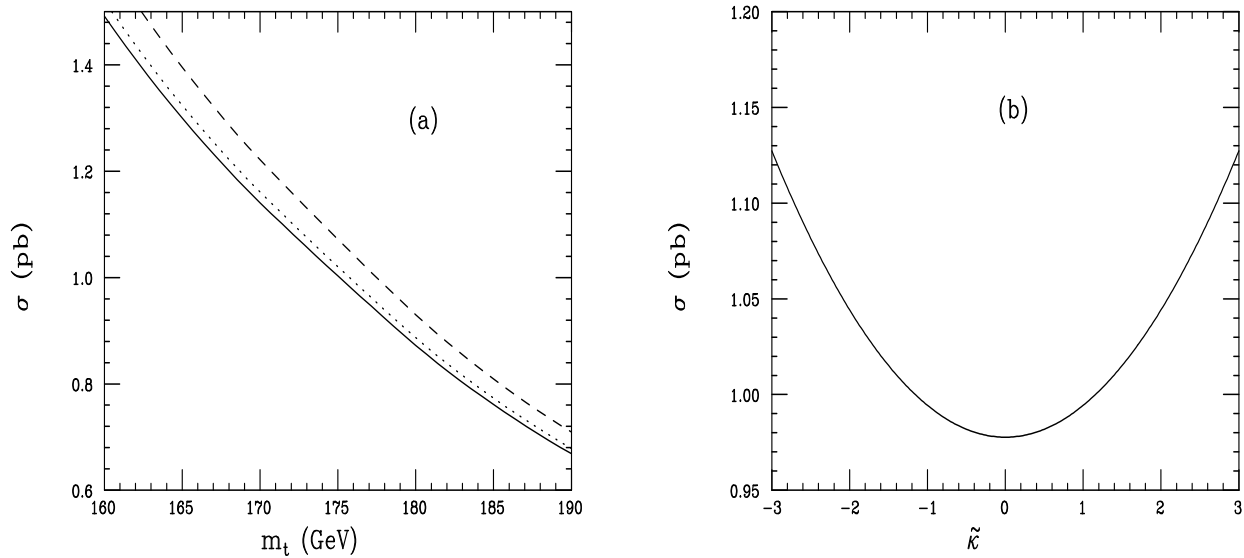

Figure 4: (a) Cross section for the process $gW^+ \to t\bar{b}(+ h.c.)$ as a function of $m_t$ at the Tevatron. The solid curve is the SM prediction whereas the dotted or dashed curve corresponds to $\tilde{\kappa} = 1$ or 2, respectively. MRSA parton densities are assumed. (b) $\tilde{\kappa}$ dependencies of the cross section shown in (a) for $m_t = 175$ GeV. In both plots, $\kappa = 0$ is assumed.

Of course we might ask if other observables are better probes of non-zero anomalous couplings than just the cross sections themselves. In our previous[4] work we showed that this was *not* the case for pair production of tops at the Tevatron due to the fact that the cross section was dominated by the region near the production threshold. What about top pairs at the LHC? Figs. 6 and 7 show the $t\bar{t}$ invariant mass $(M)$, transverse momentum$(p_t)$, rapidity$(y)$, and center of mass scattering angle$(cos\theta^*)$ distributions for the LHC for several values of $\kappa$ as well as the SM. These were obtained following the same procedure as in our previous analysis[4]. Also shown, in Figs. 6b and 6d, are the ratios of the $M$ and $p_t$ distributions to their SM values, i.e., $R_M$ and $R_{p_t}$. Although not all independent, these distributions inform us that at the LHC both the $M$ and $p_t$ distributions have comparable sensitivities to non-zero values of $\kappa$ as does the total cross section itself, i.e., values of $\kappa$ of order 0.1 will be readily separable from the SM.

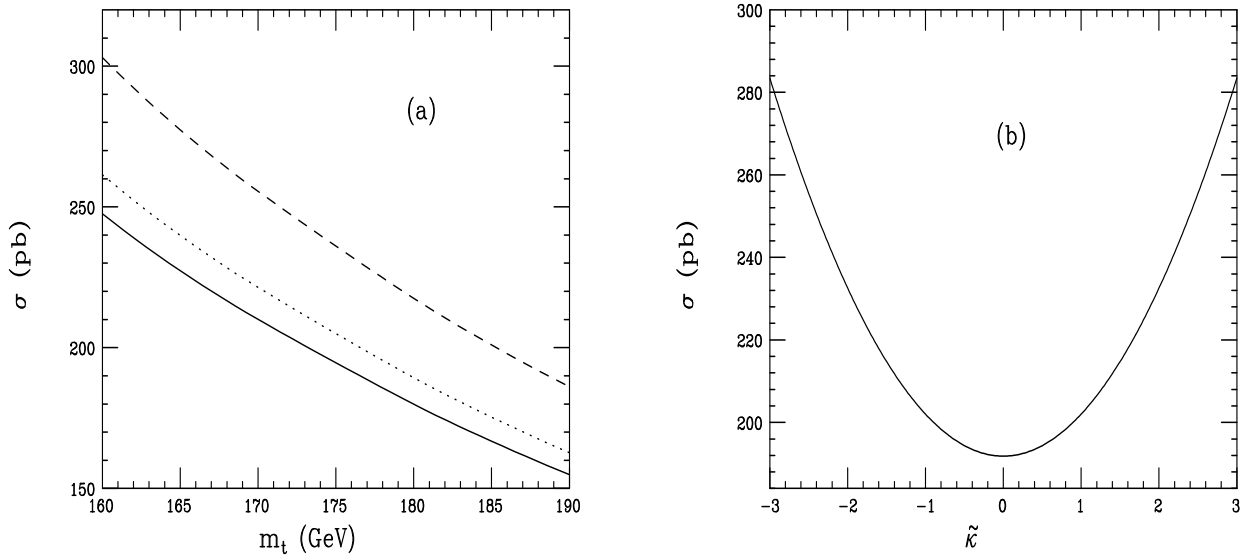

Figure 5: Same as Fig.4 but for the LHC.

Unfortunately, the same distributions for single top production at either the Tevatron or LHC do not show sensitivities to anomalous couplings even remotely comparable to what we have just seen for pair production. Fig.8 shows the $t\bar{b}$ invariant mass and $z = cos\theta^*$ distributions for single top production at the Tevatron and LHC for the SM as well as for several large values of $\kappa$. For non-zero $\tilde{\kappa}$, the results lie midway between the two curves with the corresponding values of $|\kappa|$. Even for these large values of $\kappa$ or $\tilde{\kappa}$ we see that the distributions at the Tevatron are not particularly useful as probes of anomalous couplings. The $z$ distributions are a bit more interesting, particularly at LHC energies. Note that as $z \to -1$ where the $b$—exchange dominates the amplitude all sensitivity to anomalous couplings completely vanishes, *i.e.*, all of the sensitivity comes from the 'forward' direction where the cross section is smallest. Imposing a modest angular cut at the LHC, say $z > 0$, would cleanly allow separation between the SM and $|\kappa| \simeq 1 - 2$. However, this level of sensitivity is still about an order of magnitude worse that the pair production channel.

In this paper, we have examined the single production of top quarks via $gW$ fusion assuming the existence of anomalous chromomagnetic and/or chromoelectric dipole moment $t\bar{t}g$ couplings. The analysis was performed for both the Tevatron as well as the LHC. Our main results can be summarized as follows. Since the $gW$ fusion process cross section is about a factor of 5 smaller than that for top pairs via $gg + q\bar{q}$ annihilation, a substantially larger sensitivity is needed in the $gW$ channel for it to be competitive. Unfortunately, for either chromomagnetic or chromoelectric moments we found sensitivities more than an order of magnitude smaller than in the annihilation channel from considerations of the total cross section as well as various kinematic distributions. We thus can conclude that the annihilation channel offers the best opportunity to hunt for anomalous top-gluon couplings at hadron colliders, although the single production mode may provide a cross check on the underlying physics.

## ACKNOWLEDGEMENTS

The author like to thank A. Kagan, D. Atwood, J.L. Hewett, E. Berger, D. Amidei, F. Paige, M. Hildreth and P. Burrows for discussions related to this work. He would also like to thank the members of both the Argonne National Laboratory High Energy Theory Group as well as the Phenomenology Institute at the University of Wisconsin-Madison for use of their computing facilities and kind hospitality.

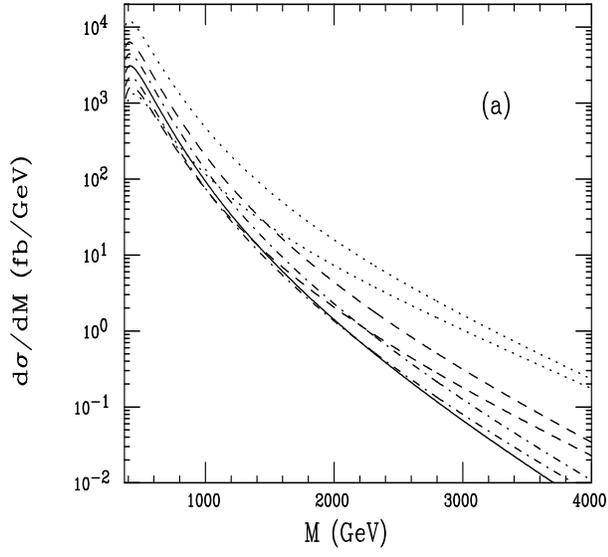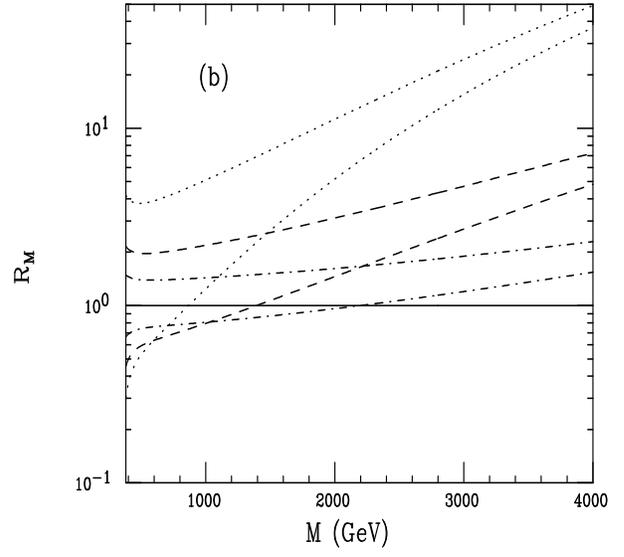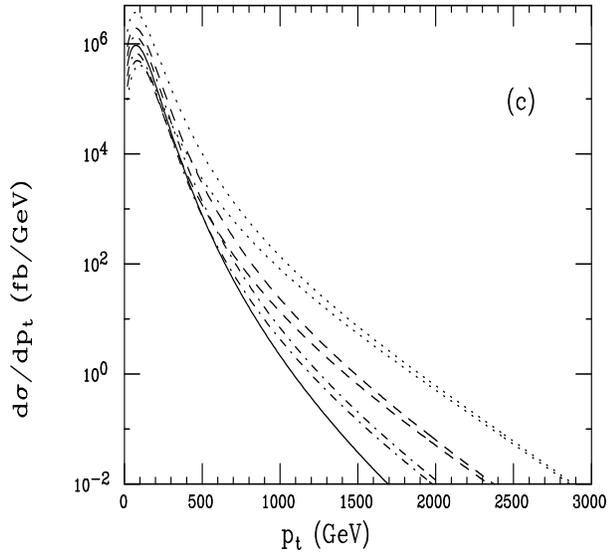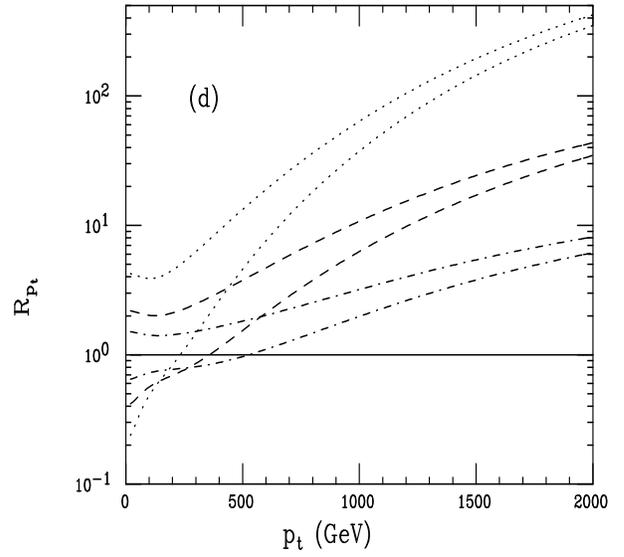

Figure 6: (a) $t\bar{t}$ invariant mass distribution at the LHC for various values of $\kappa$ assuming $m_t = 180$ GeV. (b) The same distribution scaled to the SM result. (c) $t\bar{t}$ $p_t$ distribution at the LHC and (d) the same distribution scaled to the SM. In all cases, the SM is represented by the solid curve whereas the upper(lower) pairs of dotted(dashed, dash-dotted) curves corresponds to $\kappa =$0.5(-0.5), 0.25(-0.25), and 0.125(-0.125), respectively.

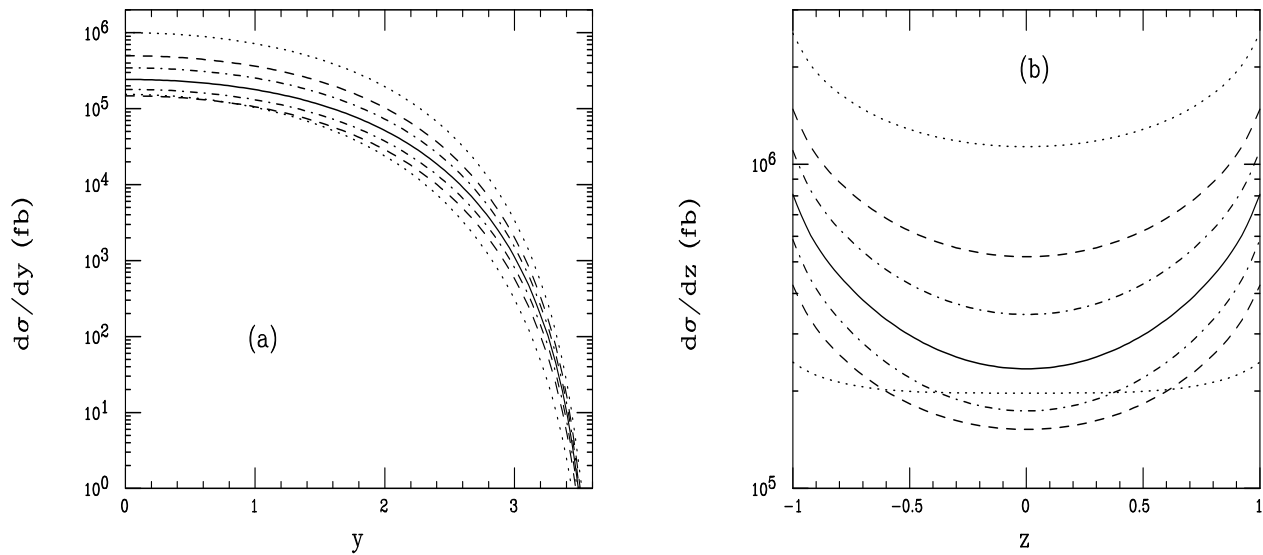

Figure 7: (a) Rapidity and (b) $z = cos\theta^*$ distributions for top quark pair production at the LHC assuming $m_t = 180$ GeV. The curves are labeled as in the previous figure.

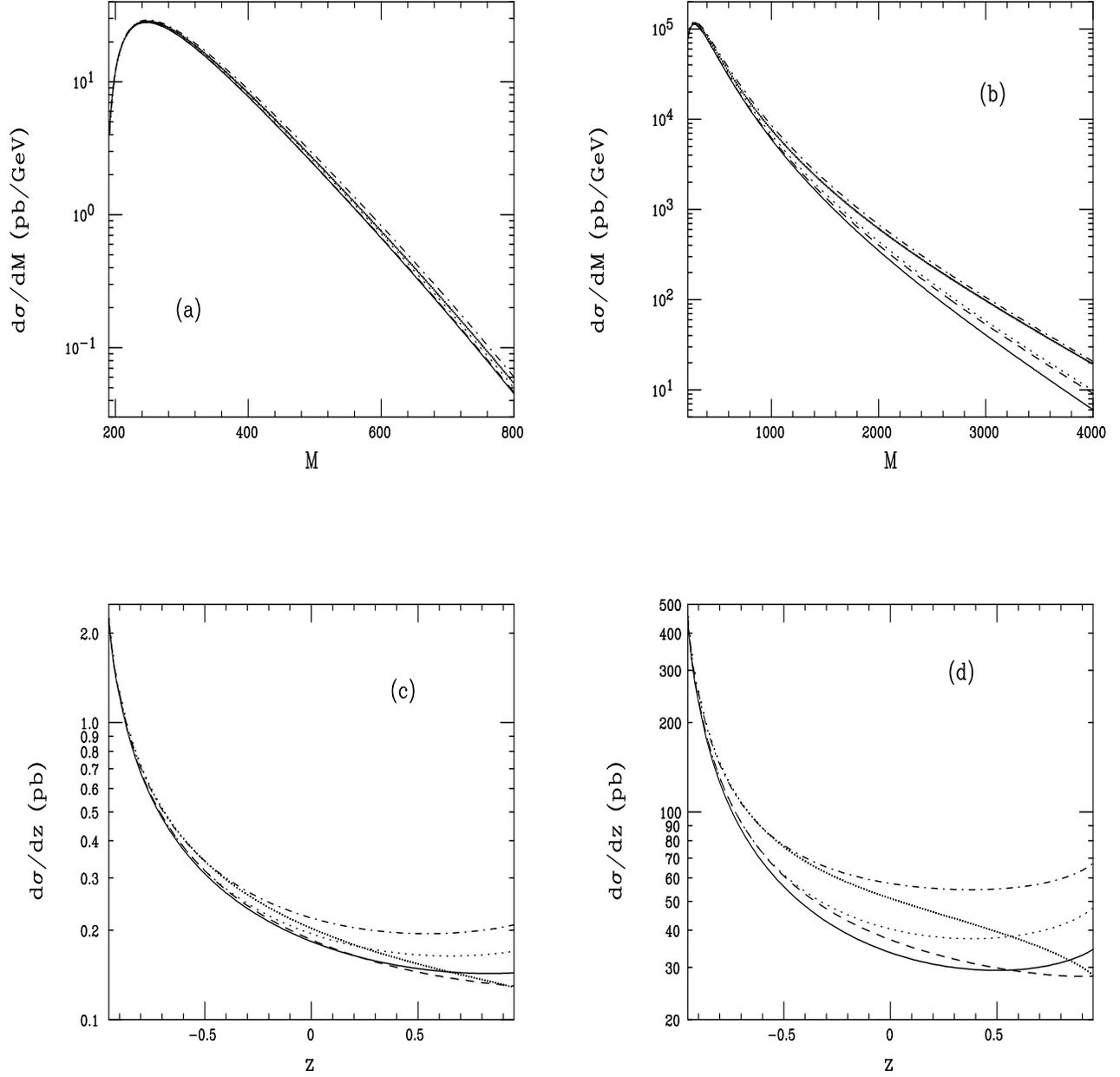

Figure 8: $t\bar{b} + b\bar{t}$ invariant mass distribution from $Wg$ fusion at (a)the Tevatron and (b)LHC. $cos\theta^*$ distribution for the same process at (c)the Tevatron and (d)LHC. In all cases the curves are labelled as in Fig.1. MRSA parton densities are assumed.

17